\documentclass[twocolumn]{aastex631}
\usepackage[utf8]{inputenc}
\DeclareUnicodeCharacter{2212}{-}
\usepackage{amsmath}
\usepackage{amssymb}
\usepackage[T1]{fontenc}
\usepackage{lmodern}
\usepackage{savesym}
\savesymbol{tablenum}
\usepackage{siunitx}
\usepackage{booktabs}
\usepackage{chemformula}
\restoresymbol{SIX}{tablenum}
\usepackage{multirow}
\usepackage{subfigure}

\graphicspath{{./}{figures/}}

\begin{document}

\title{Unveiling the interplay between the GASP jellyfish galaxy JO194 and its environment with {\it Chandra}}

\correspondingauthor{Alessandro Ignesti}
\email{alessandro.ignesti@inaf.it}

\author{Chiara Bartolini}
\affiliation{Dipartimento di Fisica e Astronomia, Università di Bologna, via Piero Gobetti 93/2, 40129 Bologna, Italy}

\author[0000-0003-1581-0092]{Alessandro Ignesti}\affiliation{INAF-Padova Astronomical Observatory, Vicolo dell’Osservatorio 5, I-35122 Padova, Italy}

\author[0000-0002-0843-3009]{Myriam Gitti}\affiliation{Dipartimento di Fisica e Astronomia, Università di Bologna, via Piero Gobetti 93/2, 40129 Bologna, Italy}\affiliation{INAF, Istituto di Radioastronomia di Bologna, via Piero Gobetti 101, 40129 Bologna, Italy}

\author[0000-0002-0843-3009]{Fabrizio Brighenti}\affiliation{Dipartimento di Fisica e Astronomia, Università di Bologna, via Piero Gobetti 93/2, 40129 Bologna, Italy}

\author[0000-0001-5840-9835]{Anna Wolter}\affiliation{INAF - Osservatorio Astronomico di Brera, via Brera, 28, 20121, Milano, Italy}

\author[0000-0002-1688-482X]{Alessia Moretti}\affiliation{INAF-Padova Astronomical Observatory, Vicolo dell’Osservatorio 5, I-35122 Padova, Italy}

\author[0000-0003-0980-1499]{Benedetta Vulcani}\affiliation{INAF-Padova Astronomical Observatory, Vicolo dell’Osservatorio 5, I-35122 Padova, Italy}

\author[0000-0001-8751-8360]{Bianca M. Poggianti}\affiliation{INAF-Padova Astronomical Observatory, Vicolo dell’Osservatorio 5, I-35122 Padova, Italy}

\author[0000-0002-7296-9780]{Marco Gullieuszik}\affiliation{INAF-Padova Astronomical Observatory, Vicolo dell’Osservatorio 5, I-35122 Padova, Italy}

\author[0000-0002-7042-1965]{Jacopo Fritz}\affiliation{Instituto de Radioastronomía y Astrofísica, UNAM, Campus Morelia, A.P. 3-72, C.P. 58089, Mexico}

\author[0000-0002-8238-9210]{Neven Tomi\v{c}i\'{c}}\affiliation{Dipartimento di Fisica e Astronomia, Università di Firenze, Via G. Sansone 1, 50019 Sesto Fiorentino, Firenze, Italy}\affiliation{INAF – Osservatorio Astrofisico di Arcetri, Largo E. Fermi 5, 50127 Firenze, Italy}

\begin{abstract}
X-ray studies of jellyfish galaxies opened a window in the physics of the interplay between intracluster medium (ICM) and interstellar medium (ISM). In this paper, we present the study of an archival \textit{Chandra} observation of the GASP jellyfish galaxy JO194. We observe X-ray emission extending from the stellar disk to the unwinding spiral arms with an average temperature of $kT=0.79\pm0.03$ keV. To investigate the origin of the X-ray emission, we compare the observed X-ray luminosities with those expected from the star formation rates (SFR) obtained from H$\alpha$ emission. We estimate an X-ray luminosity excess of a factor $\sim2-4$ with respect to the SF, therefore we conclude that SF is not the main responsible for the extended X-ray emission of JO194.  The metallicity in the spiral arms ($Z=0.24^{+0.19}_{-0.12}~\si{Z_{\odot}}$) is consistent with that of the ICM around JO194 ($Z=0.35\pm0.07$), thus we suggest that the ICM radiative cooling dominates the X-ray emission of the arms. We speculate that the X-ray plasma results from the ISM-ICM interplay, although the nature of this interplay is still mostly unknown. Finally, we observe that the X-ray properties of JO194 are consistent with those of two other GASP galaxies with different stellar mass, phase-space conditions in their hosting clusters, and local ICM conditions. We suggest that the conditions required to induce extended X-ray emission in jellyfish galaxies are established at the beginning of the stripping, and they can persist on long time scales so that galaxies in different clusters and evolutionary stages can present similar extended X-ray emission.

\end{abstract}


\section{Introduction} \label{sec:intro}

The evolution of galaxies is strongly influenced by the environment in which they live. Star-forming galaxies in high density environments, like clusters, have a different gas content than their counterparts in the fields. In particular, spiral galaxies in clusters have, on average, a lower gas content \citep[e.g.,][]{1973MNRAS.165..231D, 1988gera.book..522G, 2001ApJ...548...97S, Chung_2009} than the galaxies in the field.  This is due to a number of physical interactions that can be particularly efficient in galaxy clusters, and will eventually lead to the gas removal. As a consequence of the loss of gas, the star formation rate (SFR) is reduced and the galaxies could become passive \citep[e.g.,][]{Werle_2022}. To explain the lack of gas-rich galaxies in clusters, a mechanism that strips away the gas from galaxies was proposed by \cite{1972ApJ...176....1G}: ram pressure stripping. This mechanism acts when a galaxy falls into the hot and dense ($n_{\text{ICM}}\sim 10^{-4}-10^{-2}~\si{c\meter^{-3}}$, \citealt[][]{1986RvMP...58....1S}) intracluster medium (ICM). When a galaxy falls into the ICM, it experiences a drag force in the opposite direction of its relative motion and, if this force exceeds the gravitational one, the cold interstellar medium (ISM) of the galaxy is stripped away. The condition that expresses that the ram pressure overcomes the gravitational force per unit of surface is $\rho_{\text{ICM}}v^{2}>2\pi G\Sigma_{\ast}\Sigma_{g}$ \citep[][]{1972ApJ...176....1G}, where $\rho_{\text{ICM}}$ is the ICM mass density, $v$ is the infall velocity of the galaxy, $\Sigma_{\ast}$ and $\Sigma_{g}$ are the stellar and gas surface density. If this condition is true, the whole ISM is stripped away and the galaxy becomes passive. \\

The most spectacular examples of galaxies undergoing strong ram pressure are the so-called “jellyfish galaxies”. These galaxies show extra-planar, unilateral debris visible in the optical/UV light and long “tentacles” or tails of diffuse ionized H$\alpha$ gas extending for dozens of $\si{\kilo pc}$ beyond the galaxy disk, where new stars are born in knots \citep[e.g.][]{10.1093/mnras/stu2092, 2016AJ....151...78P}. The `GAs Stripping Phenomena in galaxies' (GASP) survey aims at  studying the gas removal process and how ram pressure stripping affects star formation \citep[e.g.,][]{2018ApJ...866L..25V, Poggianti2019,Vulcani_2020, Bellhouse2021}, and AGN activities \citep[][]{2017Natur.548..304P,Radovich_2019,Peluso_2022} in jellyfish galaxies. This project is a European Southern Observatory (ESO) Large Program, carried out with the Multi Unit Spectroscopic Explorer (MUSE) of the Very Large Telescope. The 94 stripping candidates \citep[][]{2017Natur.548..304P} show an unilateral debris/disturbed morphology and tails, they are at $z=0.04-0.07$, they have stellar masses in the range $10^{9.2}-10^{11.5}~\si{M_{\odot}}$ and they are located in different environments, such as groups and clusters with different masses \citep[][]{Poggianti_2017_b}.\\ 

With multi wavelength studies it is possible to investigate the process of ram pressure stripping through the observation of stripped gas in different phases \citep[e.g.,][]{Poggianti2019a}. With observations in the optical band it is possible to reconstruct the star formation history of the galaxies \citep[e.g.,][]{2017ApJ...848..132F} and, analyzing them in the radio band, it is possible to understand what is the role of magnetic fields during the stripping process \citep[e.g.,][]{Muller2020}. In this context, X-ray studies of these galaxies can probe the physics of the interplay between ISM and ICM. The high-energy side of jellyfish galaxies has been the object of several studies in the past decades \citep[e.g.,][]{Cowie1977, 10.1093/mnras/198.4.1007,refId0, Sun2005, 2006ApJ...637L..81S,Sun2009, Zhang2013, Boselli2016, Poggianti2019a,2021ApJ...911..144C,Sun2021}. These suggest that the observed X-ray plasma is the result of a complex interaction between ICM and ISM which causes the heating of the ISM through either shock and conduction, or the cooling of the ICM onto the tails of the galaxy, or the ICM-ISM mixing. To improve the knowledge about the interaction between ICM and ISM, it is necessary to extend these studies to additional jellyfish galaxies.\\

JO194 ($z=0.041$), located in Abell 4059, is a jellyfish galaxy characterized by an extended X-ray emission outside the disk. The galaxy is almost face-on along the line-of-sight with an inclination angle between the galaxy and its direction of motion through the ICM of $8^{\circ +16}_{-5}$ \citep[][]{Bellhouse2021}. For these reasons, it is a good candidate to study the interaction between the ICM and ISM in the stripped arms in the X-ray energy band. JO194  is a spiral Sb galaxy (RA $23:57:00.680$ Dec $-34:40: 50.10$ \citealt[][]{2009A&A...497..667V}) at a projected distance of $269~\si{\kilo pc}$ North of ESO 349-G010, the central galaxy of the cluster. JO194 has a stellar mass of $M_{\ast}=1.6\cdot 10^{11}~\si{M_{\odot}}$ and a line-of-sight velocity of $v_{los}=2030~\si{\kilo\meter/\second}$ with respect to the mean velocity of the cluster \citep[][]{2017Natur.548..304P}. Its projected vicinity to the center of the cluster and its high velocity makes the stripping process very efficient. From this information, it was argued that this is the first time that the galaxy is falling to the center of the cluster \citep[see][]{2017ApJ...844...49B, Jaffe2018}. The star formation of this galaxy \citep[$9\pm2$ M$_\odot$ yr$^{-1}$][]{2018ApJ...866L..25V} is mainly located in the galaxy disk with only the $5\%$ of the stars are currently formed in the tails \citep[][]{Poggianti2019}.
JO194 was studied in \citet[][]{Bellhouse2021} to understand the unwinding mechanism of the arms caused by the stripping process. It was shown that the youngest stars ($<20~\si{Myr}$) are born in the stripped arms while older stars ($5.7\cdot 10^{9}-1.4\cdot 10^{10}~\si{yr}$) were born in the inner part of the unwinding arms near the stellar disk. So the youngest stars born in the arms trace the motion of the unwinding arms that characterize the galaxy. This is an important result to better understand how ram pressure is acting on the arms of the spiral galaxies and to reconstruct the unwinding phenomenon as a function of the stellar age \citep[][]{Bellhouse2021}.\\

Abell 4059 is a relaxed, cool-core cluster \citep[][]{Lagana2019} at $z=0.048$ \citep[][]{Abell1989}, dominated by the cD galaxy ESO 349-G010 (RA 23: 57: 00.74 Dec -34: 45: 32.99, \citealt[][]{Huang1998}). In \cite{Schwartz1991} and \cite{Huang1998} are presented the X-ray analysis of the cluster using, respectively, High Energy Astronomy Observatory (HEAO), EXOSAT, Einstein and ROSAT observations. The previous X-ray analysis revealed that there is an excess of surface brightness in the central region of the cluster. This excess is the result of a cooling flow with a nominal cooling mass rate of $\dot{M}=184^{+22}_{-25}~\si{M_{\odot}/yr}$. The analysis of the X-ray properties of the ICM of Abell 4059 are presented in \cite{Reynolds2008}, \cite{Yun2004} and \citet[][]{Mernier_2015}. From the analysis of the radial profiles, the temperature of the ICM increases with radius (in the first $\sim 100~\si{\kilo pc}$ from the center of the cluster the temperature grows from 1 to $\sim 5~\si{\kilo eV}$), while electron density ($n_{e}$) and pressure (P) decrease: $n_{e}$ goes from $0.1~\si{c\meter^{-3}}$ to $0.001~\si{c\meter^{-3}}$ and P goes from $10^{-10}~\si{erg/ \centi\meter^{3}}$ to $10^{-11}~\si{erg/ \centi\meter^{3}}$ in the first $150~\si{\kilo pc}$ (see Section \ref{ICM_analisys}). Also the abundance of different metals ($\ch{O}, \ch{Ne}, \ch{Si}, \ch{S}, \ch{Ar}, \ch{Ca}$ and $\ch{Fe}$) decreases with radius and it is peaked toward the center of the cluster. Also ram pressure stripping might have contributed to a more recent enrichment in the core \citep[][]{Mernier_2015}.\\

We carried out an X-ray analysis of the jellyfish galaxy JO194 to explore the physical properties of the X-ray plasma in the galaxy, like average temperature, metallicity and luminosity, and to investigate the contribution of the SFR in producing this emission. The final purpose is to study the origin of the X-ray emission produced in the arms of this galaxy and what are the similarities between JO194 and the other GASP jellyfish galaxies that have been studied in the X-ray energy band, JO201 and JW100. This paper is structured as follows. In Section \ref{sec:data reduction} we describe the procedures adopted for the \textit{Chandra} data reduction, for the surface brightness analysis and the spectra extraction. In Section \ref{results} we present the results of the spectral analysis of the ICM spectra and JO194 spectra and we discuss them in Section \ref{discussion}. Summary and conclusions are presented in Section \ref{conclusion}.

Throughout this paper, we adopt a Chabrier initial mass function (IMF; \citealt{Chabrier2003}), a concordance $\Lambda$CDM cosmology of $\Omega_{\text{M}} = 0.3$, $\Omega_{\Lambda} = 0.7$, $H_{0} =70~\si{\kilo\meter \second^{-1} \mega pc^{-1}}$ ($1~\si{arcsec}=0.94~\si{\kilo pc}$ at $z=0.048$) and all the errors are given at $1\sigma$ level.

\section{Data analysis} \label{sec:data reduction}

\subsection{Data preparation}

\label{obID}
The \textit{Chandra} observation of the cluster A4059 was performed with the Advanced CCD Imaging Spectrometer (ACIS) on January 26 2005 (obsID:5785, PI Reynolds and $92.1~\si{\kilo\second}$ exposure time). The observation was made in the VFAINT mode with both ACIS-S (S1,S2, and S3 chips) and ACIS-I (I2 and I3 chips) instruments.  
In this work, we processed only the CCD including JO194 (S3 chip).\\

\begin{figure*}[ht]
    \centering
    \includegraphics[scale=0.6]{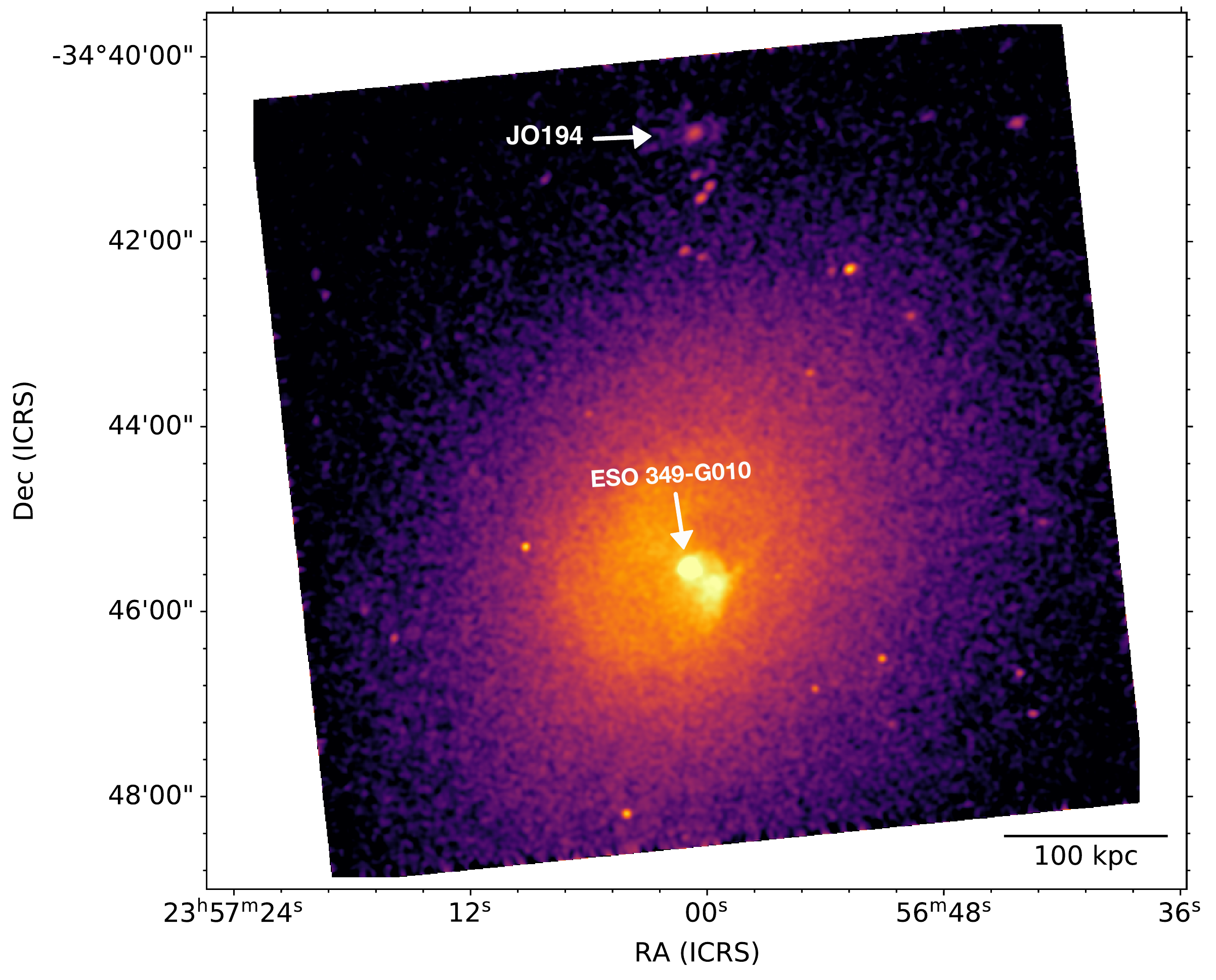}
    \caption{Exposure-corrected, background-subtracted image of Abell 4059 in the energy band $0.5-7.0~\si{\kilo eV}$ smoothed with a $3\sigma$ Gaussian.}
    \label{fig:bacground_subtracted}
\end{figure*}

The dataset was downloaded from \textit{Chandra Data Archive}\footnote{\url{https://cda.harvard.edu/chaser/}} and was reprocessed with the software package CIAO 4.13 and CALDB 4.9.4 to remove bad pixels and flares.  
After the cleaning phase, the net exposure time was $91.36~\si{\kilo\second}$.
To improve the absolute astrometry, we have identified the point sources using the CIAO task \texttt{WAVDETECT} and we cross-matched them with the optical catalog USNO-A2.0\footnote{\url{http://tdc-www.harvard.edu/catalogs/ua2.html}}.
To evaluate the background signal (observational and cosmological) we used the CALDB \texttt{Blank-Sky} files that match our data in the chip of interest. Then we normalized the observed and background count-rate in the $9-12~\si{\kilo eV}$ band rescaling the exposure time of the background accordingly. The exposure-corrected, background-subtracted \textit{Chandra} image of the chip of interest in the $0.5-7.0~\si{\kilo eV}$ energy band is shown in the Fig. \ref{fig:bacground_subtracted}. The resolution at the off-axis distance of JO194 is $4.92~\si{arcsec}.$\\

In addition to the \textit{Chandra} data, in this work we made use of the MUSE corrected for stellar absorption H$\alpha$ image of JO194 \citep[][]{2017Natur.548..304P}, and the stellar disk mask presented in \citet[][]{Gullieuszik_2020}.

\subsection{Surface brightness analysis}
\label{brightness}

To characterize the X-ray morphology of JO194, we analyzed the X-ray surface brightness profiles of the ICM of Abell 4059. We performed the analysis using the \textit{sherpa}\footnote{\url{https://cxc.cfa.harvard.edu/sherpa/threads/}} environment of the software CIAO. \\

To analyze the 2D surface brightness profile, we defined an ellipsoidal region that followed the morphology of the X-ray emission (Figure \ref{fig:bacground_subtracted}), centered on the coordinates of the cluster center (RA (J2000) 23 : 57 : 01.065, Dec (J2000) $−34 : 45 : 33.28$\footnote{These coordinates are shown in the ACCEPT archive: \\ \url{https://web.pa.msu.edu/astro/MC2/accept/clusters/5785.html}}), with major axis of $246.7~\si{\kilo pc}$ and position angle of $345^{\circ}$. To model only the ICM thermal emission, we excluded the point sources found by the task \texttt{WAVDETECT} with circles of average radius of $12~\si{arcsec}$. With \textit{sherpa}, we performed a 2D fit of the surface brightness $I(r)$ inside this ellipsoidal region using a Cash statistic \citep[][]{1979ApJ...228..939C} and an elliptical\footnote{\url{https://cxc.cfa.harvard.edu/sherpa/ahelp/beta2d.html}} $\beta$-model \citep[][]{CavaliereFusco}:
\begin{equation}
    I(r)=I_0\left[1-\left(\frac{r}{r_c} \right)^2 \right]^{0.5-3\beta}
\end{equation}

where $I_0=(3.60\pm0.02)\cdot10^{-8}~\si{photons/c\meter^{2}/pixel^{2}/s}$ is the central surface brightness, $r_c=60.53\pm0.50~\si{\kilo pc}$ is the core radius and $\beta=0.53\pm0.01$. The model's ellipticity and inclination angle are $\epsilon=0.19$ and $\theta=-5.29$ radians, respectively. The best-fit $\beta$ and $r_c$ values are in agreement with the previous results presented in \cite{Huang1998}, which reported $\beta=0.55^{+0.03}_{-0.03}$ and $r_{c}=66^{+10}_{-9}~\si{\kilo pc}$. Then, with the CIAO routine \textit{dmimgcalc}, we subtracted the 2D $\beta$-model to the exposure-corrected background-subtracted image to obtain a residual image that highlights the excesses of surface brightness of JO194 relative to the ICM thermal emission.\\

\subsection{Spectral Analisys}
\label{spectral_analisys}
To understand the origin of the extended X-ray emission of JO194 and to study the interaction between the ICM and ISM contained in the arms of the galaxy, we performed a spectroscopic analysis of the \textit{Chandra} data with the software package XSPEC (version 12.11.1 \citealt{1996ASPC..101...17A}).\\

\begin{figure*}[ht]
    \centering
    \subfigure[]{\includegraphics[scale=0.4]{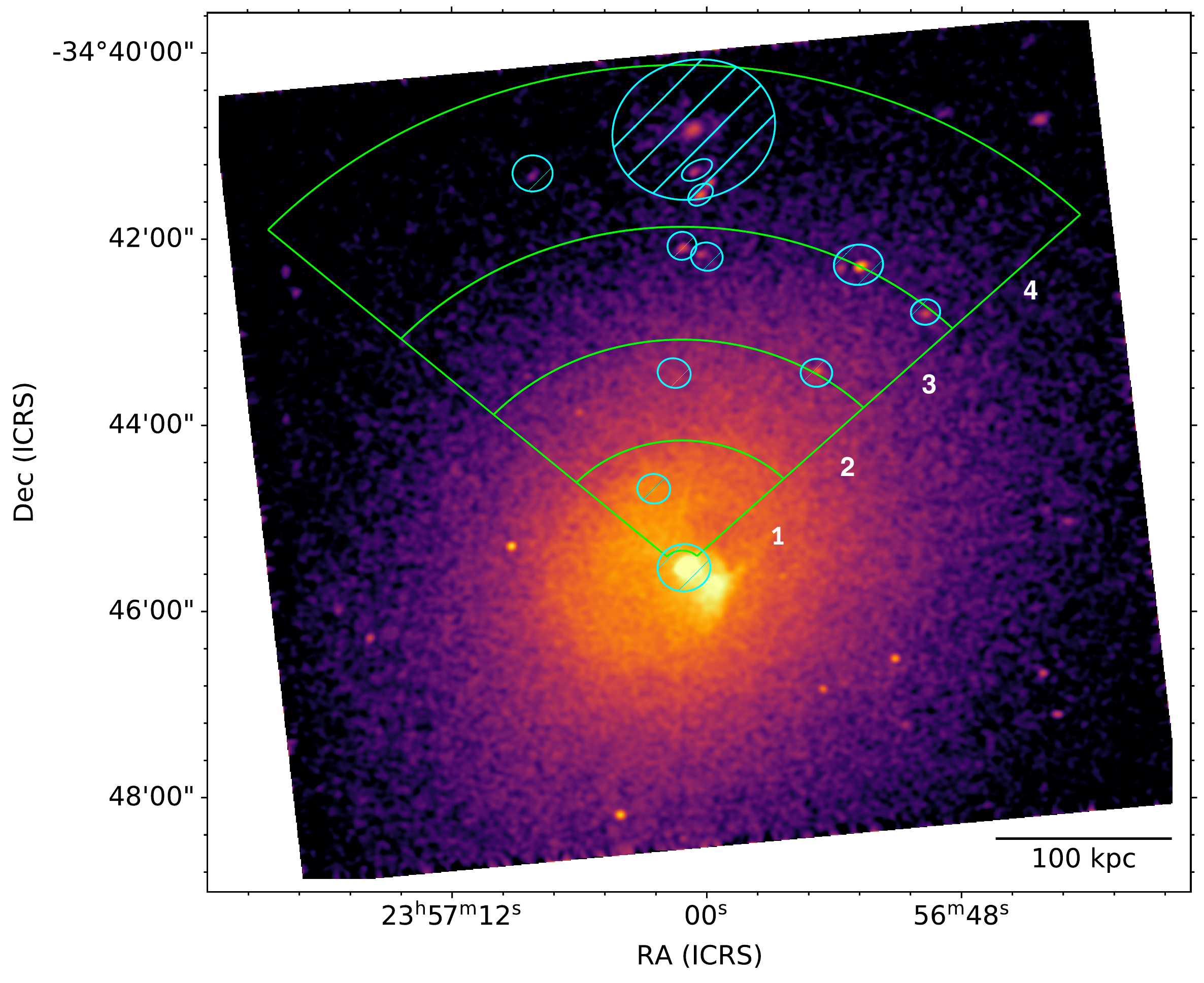}\label{fig:settori}}
    \subfigure[]{\includegraphics[scale=0.4]{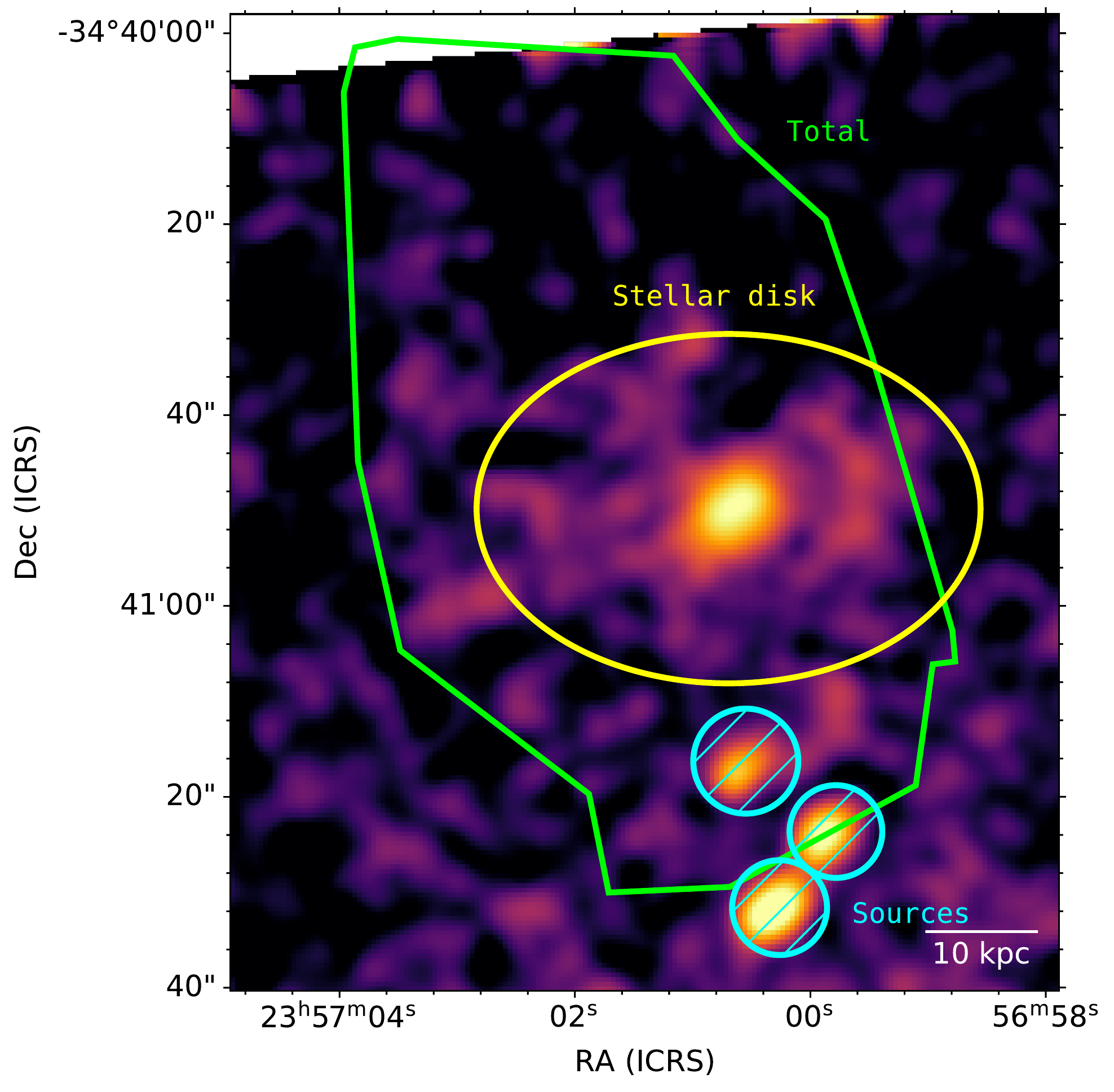}\label{fig:JO194_X_estraction}}
    \caption{(a) Exposure-corrected, background-subtracted, Gaussian-smoothed image of Abell 4059 in the $0.5-7.0~\si{\kilo eV}$ energy band. The sectors used to extract the ICM spectra (green) and the masked regions (blue) are overlaid.(b) Exposure-corrected, background-subtracted, Gaussian-smoothed image of JO194 in $0.5-7.0~\si{\kilo eV}$ energy band with on top the regions used for the spectral extraction based on the contours of the H$\alpha$ emission and the stellar disk. The three blue sources are background galaxies, according to their MUSE optical spectra. Therefore, we exclude them from the analysis.}
    \label{fig:settori_estrazione}
\end{figure*}

To derive the ICM properties in the region in which the galaxy is located, we assumed that the gas has a spherical symmetry  
and we considered 4 annular sectors with an aperture of $93^{\circ}$ centered on the center of the cluster. We adopted an average width of $77.2~\si{arcsec}=72.5~\si{\kilo pc}$ to have more than 1000 counts in every sector 
(see Figure \ref{fig:settori}). \\

For each region, we masked all the point sources identified by \texttt{WAVDETECT} and the extended emission of JO194 contained in the last annulus in the north direction, to prevent their emission from contaminating that of the ICM. We evaluated that the extended emission of JO194 is contained in an elliptical region of major axis of $45.5~\si{\kilo pc}$ and of minor axis of $41.4~\si{\kilo pc}$, shown in Figure \ref{fig:settori}.\\

For the spectral analysis of JO194, we considered different regions: a region that follows the contours of the H$\alpha$ emission \citep{2019MNRAS.482.4466P} and resembles the X-ray emission (named “Total”), a region that traces the contour of the stellar disk (named “Disk”) \citep{Gullieuszik_2020} and a region that is the difference between the Total and the Disk regions, from now on we will call this one “Arms of JO194” because it covers the unwinding arms of the galaxy.  Figure \ref{fig:JO194_X_estraction} shows the regions in which the spectra of JO194 are extracted in the $0.5-7.0~\si{\kilo eV}$ energy band. The extraction of different regions was possible thanks to the high number of counts in all regions. Table \ref{tab.conteggi1} presents the total counts of the extracted spectra of JO194. The Total and Disk regions were chosen to compare the X-ray luminosities, obtained from the spectral analysis, with those obtained from the SFR found in a previous study on the H$\alpha$ emission of JO194 \citep{2018ApJ...866L..25V} (see section \ref{SFR}).\\ 

\begin{table}
\caption{Total data counts and rate of counts that do not belong to the background counts of the spectra that represent the Total, the Disk and the Arms of JO194.}
\label{tab.conteggi1}
\begin{tabular}{lll}
\toprule
 & Total Data Counts & Rate \\
\midrule
Total & $2881$ & $84.3\%$ \\

Disk & $1461$ & $88.5\%$ \\

Arms & $1448$ & $81.7\%$ \\
\bottomrule
\end{tabular}
\end{table}

In the spectral analysis we excluded the point sources in the field of view. Concerning the three point sources located $\sim 14~\si{\kilo pc}$ south of JO194 (Figure \ref{fig:JO194_X_estraction}), the MUSE optical spectra revealed that they are background galaxies at redshift $z>0.63$, thus they were masked. \\

We extracted a spectrum from each region using the CIAO task \texttt{specextract} and we binned them to give 25 counts in each energy bin. Background spectra were extracted using \texttt{blanksky} files in each region. All spectra of the X-ray emission of JO194 and of the ICM were extracted separately and then were analyzed in the $0.5-7.0~\si{\kilo eV}$ range.

\subsection{Spectral models}
\label{models}

To take into account that the X-ray emission of the ICM along the line of sight contaminates that of JO194, we investigated the properties of the ICM surrounding the galaxy. The spectra of the four ICM regions were fitted jointly to obtain the deprojected profiles of the ICM thermal properties. We estimated deprojected quantities by using the model \texttt{projct$\cdot$(phabs$\cdot$apec}) where \texttt{projct} is the component that executes the deprojection, \texttt{phabs} represents the galactic absorption and \texttt{apec} describes the emission of a single temperature plasma\footnote{For more information consult the website: \url{https://heasarc.gsfc.nasa.gov/xanadu/xspec/manual/node133.html}}. The Galactic line-of-sight absorption was fixed to a column density value of $n_{H}=1.08\cdot 10^{20}~\si{c\meter^{-2}}$ as computed in the HI4PI survey \citep[][]{HI4PI2016}. The parameters of the \texttt{apec} component, temperature, metallicity, and normalization, were left free to vary, while the redshift was fixed to the cluster value $z=0.048$ \citep[][]{Abell1989}. The results of this analysis are presented in Section \ref{results}.

We modeled separately the spectra of the regions of JO194 (see Figure \ref{fig:JO194_X_estraction}) in order to investigate the nature and the origin of the extended X-ray emission in every region. In order to have a reliable fit of the galaxy spectra, we had to take into account the ICM contamination along the line of sight. Specifically, by rescaling the ICM spectral counts of the fourth sector in the $0.5-7.0~\si{\kilo eV}$ energy band (18500 net counts), and by comparing it with the net counts extracted on the galaxy in the same band (Table \ref{tab.conteggi1}), we estimated that the ICM composes the 77$\%$ of the total counts, and the 60$\%$ of the disk's ones. For this reason, we tested complex models made of two components: the first \texttt{apec$_{\texttt{ICM}}$} component accounts for the ICM emission along the line of sight that we modeled to the properties of the surrounding ICM ($kT=4.44\pm0.16$ keV, $Z=0.35\pm0.07$ $Z_\odot$, see Section \ref{ICM_analisys}). The second component of the model represents JO194 emission. For the galactic emission, that here after is indicated with the subscript \texttt{GAL}, we tested different models:

\begin{itemize}
    \item an \texttt{apec} component that describes the X-ray emission of a single temperature plasma and can represent the SF in the galaxy;
    \item both an \texttt{apec} and a \texttt{pow} component, this latter one describes a non-thermal emission. This model models a possible AGN emission and the contribution of unresolved High mass X-ray Binaries (HXRB);
    \item a \texttt{mkcflow}\footnote{For more information see the website:\url{https://heasarc.gsfc.nasa.gov/xanadu/xspec/manual/node193.html}}component that describes the emission of a radiative cooling multi-phase plasma from a temperature $T_{\textsc{HIGH}}$ to a temperature $T_{\textsc{low}}$ and returns the mass accretion rate ($\dot{M}$). In our analysis this model represents the ICM that cools onto the ISM;  
    \item a \texttt{cemekl}\footnote{For more information see the website: \url{https://heasarc.gsfc.nasa.gov/xanadu/xspec/manual/node149.html}} component that describes a multi-temperature plasma where the emission measure $EM=\int n_{e}n_{H} dV$ scales with the temperature of the plasma $EM\propto T^{\alpha}$ and T has a maximum, T$_{\textsc{MAX}}$. From T$_{\textsc{MAX}}$ and the index $\alpha$, it is possible to derive the mass-weighted temperature of the plasma, $T_{mw}= T_{\textsc{MAX}} \cdot(\alpha +1)/(\alpha + 2)$ (\cite{2010ApJ...708..946S}).
    
\end{itemize}

In all models we included the Galactic line-of-sight absorption using the \texttt{phabs} component with $n_{H}=1.08\cdot 10^{20}~\si{c\meter^{-2}}$ \citep[][]{HI4PI2016}. 

For the metallicities of the ISM we used two different approaches: the first one was to set free the metallicity, the second one was to fix it to specific values. With the first approach, in some cases the resulting best-fit parameters were unreliable (see Table \ref{tab.results}). In these cases, to model the ISM emission, we used the second approach and we adopted the metallicities of the ionized gas reported in \cite{Franchetto_2020} and the metallicity of the ICM ($Z=0.35\pm0.07~\si{Z_{\odot}}$). For the Total and Disk regions, two estimates of the oxygen abundance are provided based on different estimators: $Z_{\text{O3N2}}$, which is based on the O3N2 indicator, and $Z_{\text{PYQZ}}$ which was obtained with the PYQZ code \citep[e.g.,][]{Dopita_2013,10.1093/mnras/stv749}. The corresponding solar metallicities were derived by the following relation:
\begin{equation}
\label{metall}
    Z=10^{(12+\log O/H)-8.69}~\si{Z_{\odot}}
\end{equation}
where $8.69$ is the oxygen abundance in the sun \citep[][]{2009ARA&A..47..481A}. The metallicities values obtained for the Total region are presented in the Table \ref{tab:metallicities}.

\begin{table}[]
    \centering
    \begin{tabular}{c c c}
    \toprule
        Metallicity  & Total & Disk  \\
        \midrule
    $Z_{\text{O3N2}}$ & $1.24~\si{Z_{\odot}}$ & $1.26~\si{Z_{\odot}}$ \\
    $Z_{\text{PYQZ}}$ & $2.81~\si{Z_{\odot}}$ & $2.90~\si{Z_{\odot}}$ \\
\bottomrule    
    \end{tabular}
    \caption{Metallicities of the Total and Disk regions obtained from the results presented by \citet[][]{Franchetto_2020} for JO194 using the equation \ref{metall}.}
    \label{tab:metallicities}
\end{table}

Finally, we used C-statistic and the abundance tables presented in \cite{2009ARA&A..47..481A}.

\section{Results}
\label{results}
\begin{figure*}[ht]
    \centering
    \subfigure[]{\includegraphics[scale=0.30]{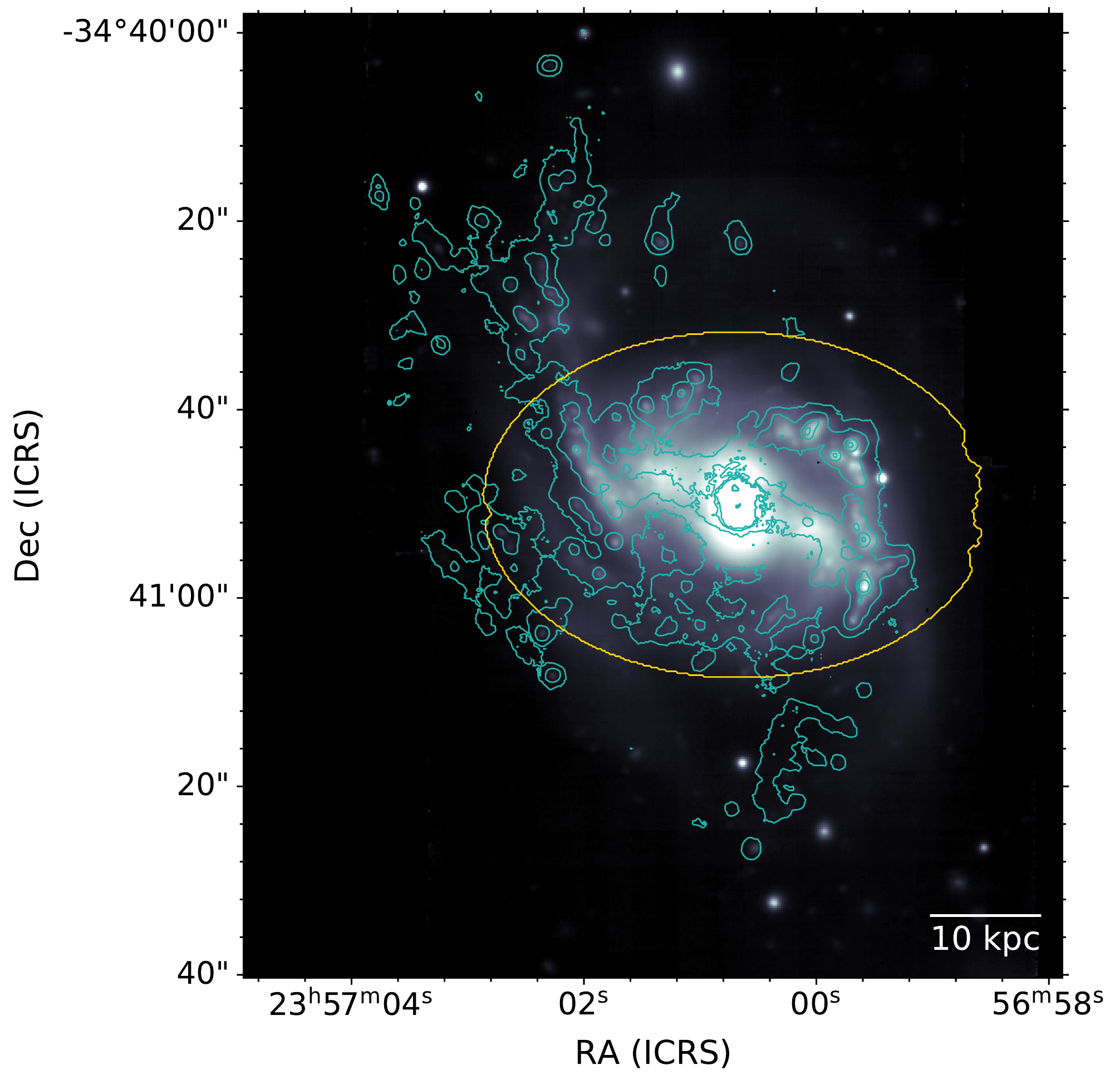}\label{fig:ottico}}
    \hspace{-0.3cm}
    \subfigure[]{\includegraphics[scale=0.30]{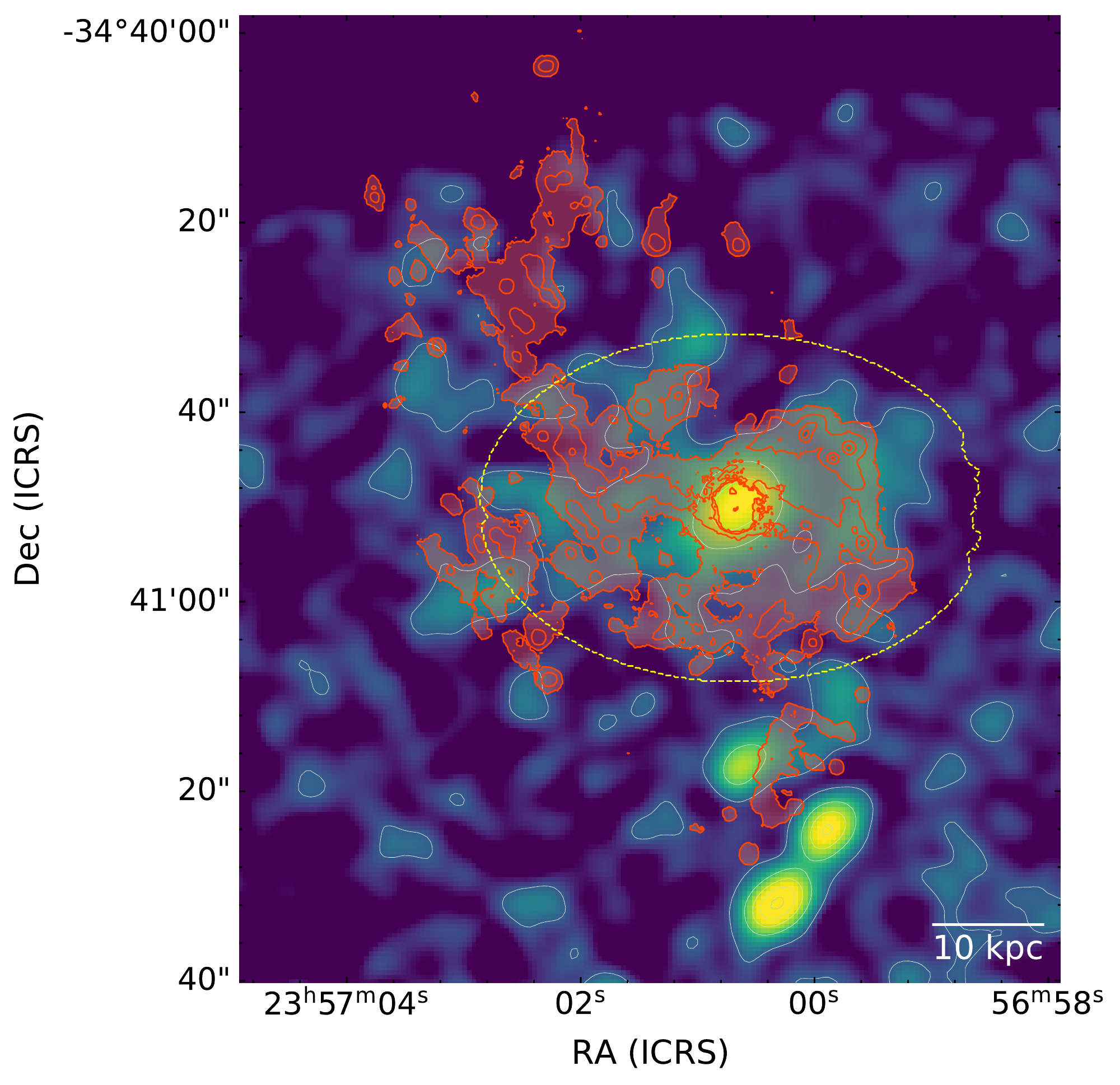}\label{residual}}
    \hspace{-0.3cm}
    \subfigure[]{\includegraphics[scale=0.30]{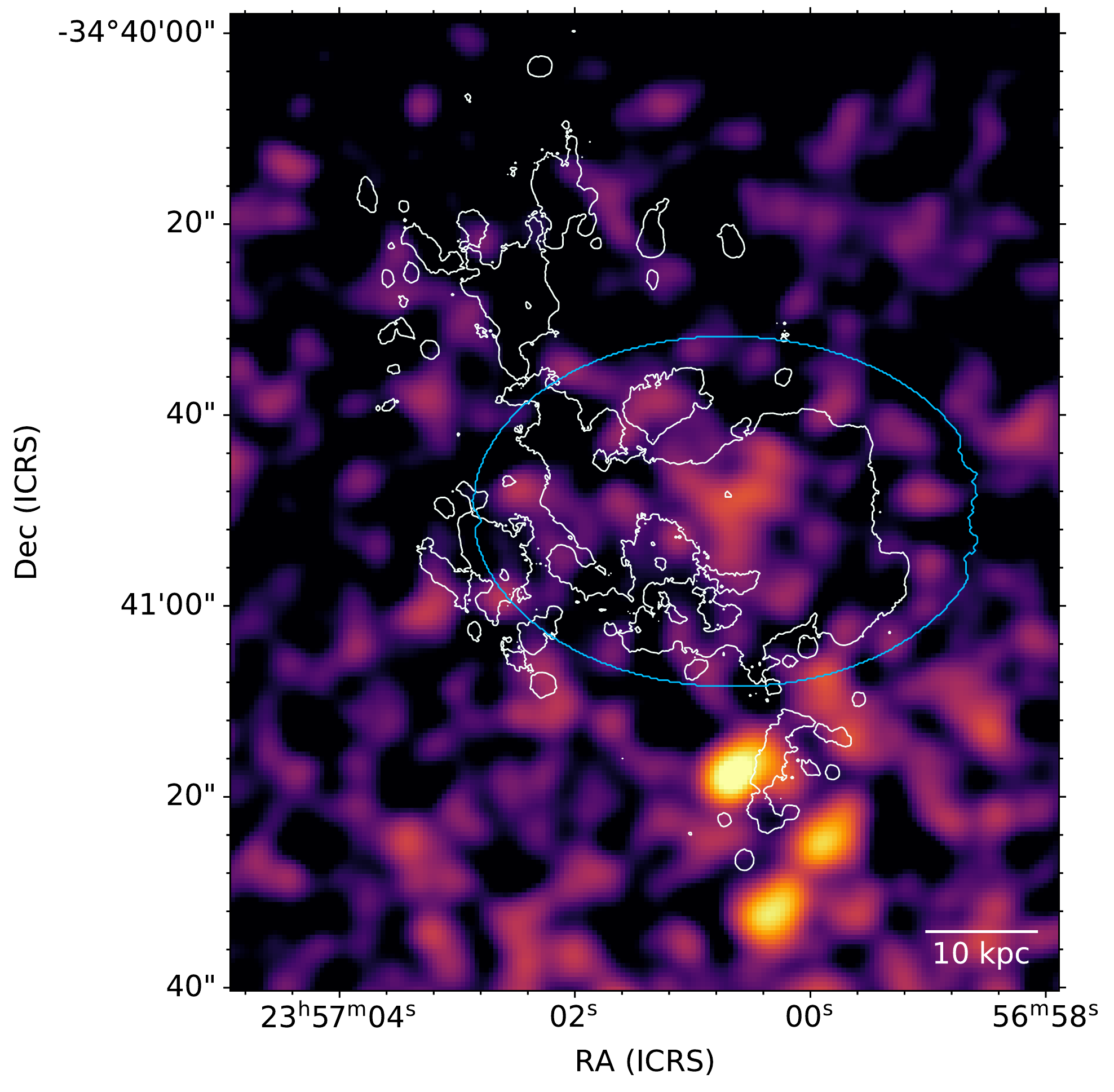}\label{hardX}}
    \caption{Multi-wavelength view of JO194 in a $1.5'\times1.7'$ region centered on the galaxy. (a) MUSE white-light image ($4650-9300~\si{\r{A}}$) with on top the contours of the stellar disk (yellow) and H$\alpha$ emission (light blue), (b)Residual X-ray image of JO194 in the $0.5-7.0~\si{keV}$ band (see Section \ref{brightness}) with on top the contours of the stellar disk (yellow) H$\alpha$ (red) and positive residuals (white); (c) Exposure-corrected, background-subtracted image in the $2.0-7.0~\si{\kilo eV}$ band, smoothed with a $3\sigma$ pixel gaussian. For reference, we show the stellar disk (blue) and the lowest contour of the H$\alpha$ emission (white).}
    \label{fig:bande}
\end{figure*}

In this section, we describe the main results of the spectral analysis of the X-ray emission of the ICM of Abell 4059 and the ISM of JO194. Figure \ref{fig:bande} shows a multi-wavelength field of JO194, in the optical, H$\alpha$ (only contours) and X-ray energy bands.

\subsection{ICM spectral analysis results}
\label{ICM_analisys}
We have first estimated the properties of the ICM around JO194. The parameters that were returned from the spectral analysis were the temperature and the metallicity of the plasma. From these, we have also estimated the deprojected electron density and pressure through the following relations: 
\begin{equation}
    n_{e}=\sqrt{\frac{N\cdot 4\pi D_{L}^{2}}{V\cdot 0.82 \cdot 10^{-14}}} 
\end{equation}
and
\begin{equation}
    P=1.9 \cdot n_{e} \cdot kT
\end{equation}

in which $N$ is the \texttt{apec} component's normalisation, $D_{L}=6.58\cdot 10^{26}~\si{c\meter}$ is the luminosity distance, $V=\frac{4}{3}\pi (R_{\textsc{ext}}^{3}-R_{\textsc{int}}^{3})$ is the volume of the spherical shell where $R_{\textsc{ext}}$ and $R_{\textsc{int}}$ are the sectors' external and internal radii, $n_{e}n_{H}=0.82n_{e}^{2}$ and the total density of particles is $n=1.9n_{e}$.\\

\begin{figure}[ht]
    \hspace{-0.2cm}
    \includegraphics[scale=0.49]{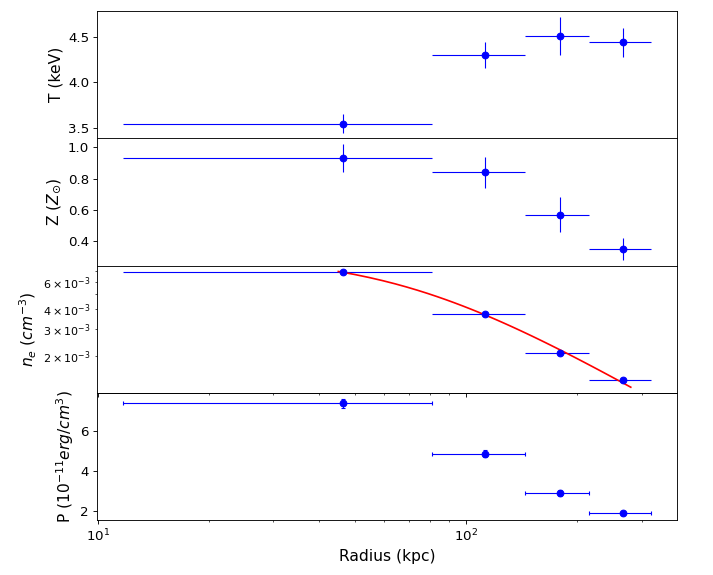}
    \caption{Deprojected radial profiles of (from top to bottom): temperature, metallicity, fitted electron density and pressure in log scale. In red we show the best-fit $\beta$-model profile ($\beta=0.46\pm0.06$ and $r_{c}=70\pm 13~\si{\kilo pc}$) which describe the deproject electron density radial profile.}
    \label{fig:profili_radiali}
\end{figure}

In Table \ref{tab.3} we report the deprojected properties of the ICM for every annulus. In Figure \ref{fig:profili_radiali} the deprojected radial profiles of the properties of the ICM are shown.
The electron density profile was fitted with a $\beta$-model to constrain the deprojected ICM density profile. The best-fit parameters are: $\beta=0.46\pm 0.06$ and core radius $r_{c}=74\pm 14~\si{arcsec}= 70\pm 13~\si{\kilo pc}$. These values are consistent with those reported in Section \ref{brightness}.

\begin{table*}[]
\caption{Internal Radius ($R_{\text{INT}}$), External Radius ($R_{\text{EXT}}$), Deprojected Temperature, Metallicity, Electronic density and Pressure of the four ICM annuli.}
\label{tab.3}
\centering
\begin{tabular}{lllll}
\toprule
Property & Annulus 1 & Annulus 2 & Annulus 3 & Annulus 4 \\
\midrule
\multirow{2}{7em}{$R_{\text{INT}}$ - $R_{\text{EXT}}$}& $12.0-82.9~\si{arcsec}$ & $82.9-148.0~\si{arcsec}$ & $148.0-220.7~\si{arcsec}$ & $220.7-325.0~\si{arcsec}$\\
& $11.3-77.9~\si{\kilo pc}$ & $77.9-139.1~\si{\kilo pc}$ & $139.1-207.4~\si{\kilo pc}$& $207.4-305.5~\si{\kilo pc}$\\

kT(keV) & $3.55\pm 0.10$ & $4.30 \pm 0.14$ & $4.51\pm 0.21$ & $4.44\pm 0.16$ \\

Z($Z_{\odot}$) & $0.93\pm 0.09$ & $0.84\pm 0.10$ & $0.57 \pm 0.11$ & $0.35 \pm 0.07$ \\

$n_{e}(10^{-3}~\si{c\meter^{-3}})$ & $6.87\pm 0.07$ & $3.74\pm 0.04$ & $2.10\pm 0.03$ & $1.39\pm 0.01$ \\

P($10^{-11}~\si{erg/c\meter^{-3}}$) &  $7.42\pm 0.25$ & $4.88\pm 0.17$ & $2.88\pm 0.15$ & $1.87\pm 0.07$ \\ 
\bottomrule
\end{tabular}
\end{table*}

\subsection{Galaxy spectral analysis results}
\label{spec_results}

As previously introduced in Section \ref{models}, we modeled the spectrum extracted over the galaxy as the combination of the ICM, represented by an {\tt apec$_{\texttt{ICM}}$} component, and galactic emissions, indicated by the subscript \texttt{GAL}. Therefore the temperature, the metallicity and the redshift parameters of the first \texttt{apec$_{\texttt{ICM}}$} component were set to those measured in the region including JO194 (i.e., the fourth sector of the ICM analysis) that are $kT=4.44\pm 0.16~\si{keV}$, $Z=0.35 \pm 0.07~\si{Z_{\odot}}$, and $z=0.048$ (see Table \ref{tab.3}). The second component of the model, that is the galactic emission, was tested among all the components shown in the list in Section \ref{spectral_analisys}. \\

The best-fit results are reported in Table \ref{tab.results}. We selected the best-fitting model based on the final statistics (i.e., best agreement with the data) and the physical significance (i.e., we rejected those models which provided a good fit with unphysical parameters, such as a temperature significantly higher than the one of the ICM, or an extremely low metallicity). In summary, we observe that: 
\begin{itemize}
    \item The “Total” spectrum is well reproduced by two different models. The first one is the \texttt{phabs$\cdot$(apec$_{\texttt{ICM}}$+cemekl$_{\texttt{GAL}}$)} with the \texttt{cemekl$_{\texttt{GAL}}$} maximum temperature fixed to the ICM temperature $T_{\textsc{MAX}}=4.44~\si{\kilo eV}$ and metallicity fixed to the values shown in Section \ref{models}. This model yields a mass weighted temperature of $T_{mw}=2.65~\si{\kilo eV}$. The second model is \texttt{phabs$\cdot$(apec$_{\texttt{ICM}}$+apec$_{\texttt{GAL}}$)} that measured an average X-ray temperature of the galactic component of  $kT=0.79~\si{\kilo eV}$, which is consistent with those of other jellyfish galaxies (see Section \ref{comparison});
    \item The best-fitting models for the galactic disk are \texttt{phabs$\cdot$(apec$_{\texttt{ICM}}$+apec$_{\texttt{GAL}}$)} with the second \texttt{apec$_{\texttt{GAL}}$} metallicity fixed to the values shown in Section \ref{models} and the model \texttt{phabs$\cdot$(apec$_{\texttt{GAL}}$+pow$_{\texttt{GAL}}$)} with the metallicity of the \texttt{apec$_{\texttt{GAL}}$} component fixed to the values of the stellar disk. Due to the low statistics, the Disk spectrum showed large uncertainties above 2 keV that resulted in small values of the reduced chi-square ($\chi_R^2\sim0.7$). In the \texttt{phabs$\cdot$(apec$_{\texttt{GAL}}$+pow$_{\texttt{GAL}}$)} model the ICM emission contribution seems to be negligible, but this is at odds with the previous results because the ICM contributes far more than the 60$\%$ of the emission in the disk (see Section \ref{models}). In addition, the resulting luminosity of the {\tt pow}$_{\tt {GAL}}$ component is not consistent with that we expect from the current SFR. By adopting the $L_{X}$-SFR relation for point sources, in the $0.5-8.0~\si{\kilo eV}$ energy band, presented in \citet[][]{2012MNRAS.419.2095M} ($L_{X, \text{non thermal}}(\si{erg/\second})=2.6\times10^{39} \cdot SFR(\si{M_{\odot}/yr})$), the current SFR in the disk of JO194 \citep[$8\pm2$ M$_\odot$ yr$^{-1}$][]{2018ApJ...866L..25V} entails a luminosity of $2.3\times10^{40}$ erg s$^{-1}$, that is more than an order of magnitude lower than we measured from the spectral analysis. Therefore the {\tt pow}$_{\tt {GAL}}$ model is not a reliable description of the disk emission, and we dismiss this model in the following analysis. We argue that the {\tt pow}$_{\tt {GAL}}$ model fit goodness turns out solely from the fact that, due to the poor statistics of the current data, it tentatively fits the spectrum above $2~\si{\kilo eV}$.
  \item The spectrum of JO194 arms is well reproduced by two different models. The first one is the \texttt{phabs}$\cdot$(\texttt{apec$_{\texttt{ICM}}$+apec$_{\texttt{GAL}}$}) model with the metallicity of the two \texttt{apec} component fixed to $Z=0.35~\si{Z_{\odot}}$. This model provides an estimate of the average temperature of arms' X-ray plasma ($kT=0.70\pm0.03~\si{\kilo eV}$). The second one is the \texttt{phabs$\cdot$(apec$_{\texttt{ICM}}$+mkcflow$_{\texttt{GAL}}$)} model with $t_{\textsc{HIGH}}$ fixed to the value of the surrounding ICM ($kT=4.44$ keV). The best-fit model yields a mass flow of $0.24^{+0.06}_{-0.05}$ $M_\odot$ yr$^{-1}$ and a metallicity of $Z=0.24^{+0.19}_{-0.12}Z_{\odot}$, which may be the first direct measure of the metallicity of the X-ray plasma in the arms of a jellyfish galaxy. The implications of these results are discussed in Section \ref{discussion}. 
\end{itemize}

\begin{table*}[t!]
\caption{Best fit results extracted from the region that follows the contours of the H$\alpha$ emission and resembles the X-ray emission (“Total”), the disk (“Disk”) and all the galaxy emission without the disk (“Arms”). The first \texttt{apec$_{\texttt{ICM}}$} component of the models refers to the ICM emission while the second refers to the galactic emission. The only exception is the fourth model in which the \texttt{apec$_{\texttt{GAL}}$} component is referred to the emission of the Disk. The best-fit parameters shown in the fourth column refer only to the galactic component (namely \texttt{apec$_{\texttt{GAL}}$}, \texttt{cemekl$_{\texttt{GAL}}$}, \texttt{pow$_{\texttt{GAL}}$} and \texttt{mkcflow$_{\texttt{GAL}}$}). The degree of freedom are indicated with the word “dof”.}
\label{tab.results}
\centering
\begin{tabular}{lllll}
\toprule
\multirow{3}{3em}{Region} & \multirow{3}{12em}{Model} & \multirow{3}{13em}{Fixed Parameters} & \multirow{3}{13em}{Parameters of the galactic component} & $\chi^{2}$\\
& & & & dof \\
& & & & $\chi^{2}_{R}$ \\
\midrule
\multirow{12}{3em}{Total} & \multirow{6}{12em}{\texttt{phabs$\cdot$(apec$_{\texttt{ICM}}$+apec$_{\texttt{GAL}}$)}} &  \multirow{3}{13em}{$Z_{\text{O3N2}}=1.24Z_{\odot}$} & $kT=0.79^{+0.03}_{-0.04}~\si{\kilo eV}$ & $\chi^{2}=77.98$ \\
& & & $F=(2.72\pm 0.22)\cdot10^{-14}~\si{erg/cm^{2}/s}$ & $dof=86$\\
& & & $L=(1.49\pm 0.12)\cdot 10^{41}\si{erg/\second}$ & $\chi^{2}_{R}=1.15$\\
\cline{3-5}

&  &  \multirow{3}{13em}{$Z_{\text{PYQZ}}=2.81Z_{\odot}$} & $kT=0.79^{+0.03}_{-0.03}~\si{\kilo eV}$& $\chi^{2}=78.35$\\
& & & $F=(2.60\pm 0.22)\cdot 10^{-14}~\si{ergs/cm^{2}/s}$& $dof=86$ \\
& & & $L=(1.42\pm 0.12)\cdot 10^{41}\si{erg/\second}$ & $\chi^{2}_{R}=0.91$\\
\cline{3-5}
 & \multirow{6}{12em}{\texttt{phabs$\cdot$(apec$_{\texttt{ICM}}$+cemekl$_{\texttt{GAL}}$)}}  & \multirow{3}{13em}{$Z_{\text{O3N2}}=1.24Z_{\odot}, T_{max}=4.44\si{\kilo eV}$} & $\alpha=0.48_{-0.18}^{+0.18}$, & $\chi^{2}=85.12$, \\
& & & $F=(5.76\pm 2.04)\cdot 10^{-14}\si{erg/c\meter^{2}/\second}$ &  $dof=85$, \\
& & & $L=(3.17\pm 1.11)\cdot 10^{41}\si{erg/\second}$ & $\chi^{2}_{R}=1.00$ \\

\cline{3-5}
& & \multirow{3}{13em}{$Z_{\text{PYQZ}}=2.81Z_{\odot}, T_{max}=4.44\si{\kilo eV}$} & $\alpha=0.46_{-0.17}^{+0.17}$ & $\chi^{2}=83.36$, \\
& & & $F=(4.80\pm 0.17)\cdot 10^{-14}\si{erg/c\meter^{2}/\second}$ & $dof=85$, \\
& & & $L=(2.64\pm 0.90)\cdot 10^{41}\si{erg/\second}$ & $\chi^{2}_{R}=0.98$ \\

\hline

\multirow{12}{3em}{Disk} & \multirow{6}{12em}{\texttt{phabs$\cdot$(apec$_{\texttt{ICM}}$+apec$_{\texttt{GAL}}$)}} & \multirow{3}{13em}{$Z_{\text{O3N2}}=1.26Z_{\odot}$} & $kT=0.85^{+0.04}_{-0.04}\si{\kilo eV}$ & $\chi^{2}=29.77$, \\
& & & $F=(1.86\pm 0.19)\cdot 10^{-14}\si{erg/c\meter^{2}/\second}$, & $dof=43$, \\
& & & $L=(1.02\pm 0.10)\cdot 10^{41} \si{erg/\second}$ & $\chi^{2}_{R}=0.69$ \\
\cline{3-5}
& & \multirow{3}{13em}{$Z_{\text{PYQZ}}=2.90Z_{\odot}$} & $kT=0.84^{+0.04}_{-0.04}~\si{\kilo eV}$ & $\chi^{2}=30.01$, \\
& & & $F=(1.77\pm 0.18)\cdot 10^{-14}\si{erg/c\meter^{2}/\second}$, & $dof=43$, \\
& & & $L=(9.70\pm 0.97)\cdot 10^{40} \si{erg/\second}$ & $\chi^{2}_{R}=0.70$ \\
\cline{3-5}
& \multirow{6}{12em}{\texttt{phabs$\cdot$(apec$_{\texttt{GAL}}$+pow$_{\texttt{GAL}}$)}} & \multirow{3}{13em}{$Z_{\text{PYQZ}}=1.26Z_{\odot}$} & $kT=0.90^{+0.04}_{-0.05}\si{\kilo eV}$, $\Gamma=1.77^{+0.10}_{-0.10}$ & $\chi^{2}=30.30$\\
& & & $F=(1.01\pm 0.18)\cdot 10^{-13}\si{erg/c\meter^{2}/\second}$ & $dof=42$\\
& & & $L=(5.48\pm0.99)\cdot10^{41}~\si{erg/\second}$ & $\chi^{2}_{R}=0.72$\\
\cline{3-5}
& & \multirow{3}{13em}{$Z_{\text{PYQZ}}=2.90Z_{\odot}$} & $kT=0.90^{+0.04}_{-0.05}\si{\kilo eV}$, $\Gamma=1.80^{+0.10}_{-0.10}$ & $\chi^{2}=30.36$\\
& & & $F=(1.01\pm 0.08)\cdot 10^{-13}\si{erg/c\meter^{2}/\second}$ & $dof=42$\\
& & & $L=(5.46\pm 0.45)\cdot 10^{41} \si{erg/\second}$ & $\chi^{2}_{R}=0.72$\\
\hline
\multirow{7}{3em}{Arms} & \multirow{3}{12em}{\texttt{phabs$\cdot$(apec$_{\texttt{ICM}}$+apec$_{\texttt{GAL}}$)}} &  \multirow{3}{13em}{$Z=0.35Z_{\odot}$} & $kT=0.70^{+0.08}_{-0.10}~\si{\kilo eV}$ & $\chi^{2}=49.21$\\
& & & $F=(9.18\cdot1.68)\cdot 10^{-15}\si{erg/c\meter^{2}/\second}$ & $dof=47$\\
& & & $L=(5.07\pm 0.92)\cdot 10^{40}\si{erg/\second}$ & $\chi^{2}_{R}=1.05$\\
\cline{3-5}
& \multirow{4}{12em}{\texttt{phabs$\cdot$(apec$_{\texttt{ICM}}$+mkcflow$_{\texttt{GAL}}$)}} & \multirow{4}{13em}{$T_{high}=4.44\si{\kilo eV}$} & $T_{low}=0.17^{+0.23}_{-0.14}\si{\kilo eV}$, $Z=0.24^{+0.19}_{-0.12}Z_{\odot}$ & \multirow{2}{5em}{$\chi^{2}=56.56$,} \\
& & & $\dot{M}=0.24^{+0.06}_{-0.05}\si{M_{\odot}/yr}$ & \multirow{2}{5em}{$dof=46$,} \\
& & & $F=(3.91\pm 0.91)\cdot 10^{-14}\si{erg/c\meter^{2}/\second}$ & \multirow{2}{5em}{$\chi^{2}_{R}=1.23$}  \\
& & & $L=(2.15\pm 0.49)\cdot 10^{41} \si{erg/\second}$ &  \\
\bottomrule
\end{tabular}
\vspace{1ex}

{\raggedright \textbf{Notes}: kT is the average temperature of the galactic plasma, $\alpha$ is the index of the powerlaw that describes how the emission measure scales with the temperature of the galaxy, $\Gamma$ is the photon index of the power law, $T_{\text{LOW}}$ is the lowest temperature of the galactic plasma and $\dot{M}$ is the mass accretion rate. The fluxes and the luminosities refer to the galactic component and were measured in $0.5-10.0~\si{\kilo eV}$ energy band.\par}
\end{table*}

\section{Discussion}
\label{discussion}

\subsection{Characteristics of the galaxy at different wavelengths}
\label{bande}

From the analysis of JO194 in the optical and H$\alpha$ energy bands, it was argued that in the center of the galaxy is located a \textit{Low Ionization Nuclear Emission-line Region} (LINER) AGN \citep[][]{2017Natur.548..304P}. From our analysis, we concluded that in the $2.0-7.0~\si{\kilo eV}$ energy band (see Figure \ref{hardX}) there is no hard X-ray emission in the center of JO194. For reference, we derive an upper limit of its luminosity in the 0.5-10 keV band of $L<6\times10^{40}$ erg s$^{-1}$. This evidence challenges the hypothesis of an AGN sitting at the center of JO194, as already discussed in \citet[][]{Radovich_2019}. For the conclusive analysis of the core region of JO194, we refer to Moretti et al., in preparation.\\

Concerning the disk in the three different energy bands, an arc of emission is visible on the right side of the disk due to an intense star formation (see Figure \ref{fig:bande}). \\

Figure \ref{fig:bande} shows the comparison between the morphology of JO194 in the optical, H$\alpha$ (only contours) and X-ray energy band. JO194 is characterized by an extended emission outside the stellar disk visible in each of these energy bands. In Figure \ref{residual}, we used the residual image of the galaxy. The morphology of the galaxy is more distinguishable in the residual image because the surface brightness excess relative to the ICM emission is highlighted. Figure \ref{hardX} shows the hard X-ray emission ($2.0-7.0~\si{\kilo eV}$) of the galaxy for the analysis of the core region of JO194. This image was used to test if there is an excess of hard X-ray emission due to the AGN. The lack of the emission at the center of this galaxy challenges the hypothesis of the presence of the AGN.

\subsection{Origin of the X-ray emission}
\label{SFR}

 The X-ray extended emission of late-type, star-forming galaxies is produced by the thermal emission of gas heated by young, massive stars, and the non-thermal X-ray emission of High Mass X-ray Binaries (HMXB) \citep[e.g.][]{Mineo_2014}. To investigate if this is the case also in JO194, we compared the observed X-ray luminosities with those expected from the ongoing star formation rates (SFR) measured in \cite{2018ApJ...866L..25V}. In \cite{2018ApJ...866L..25V} the SFR were estimated from the H$\alpha$ emission using the relation SFR ($M_{\odot}$ yr$^{-1}$) = $4.6\cdot10^{-42}(L_{H\alpha}/(\text{erg s}^{-1}))$. The SFR values for the disk and all the entire galaxy are SFR$_{\textsc{tot}} = (9\pm 2)~\si{M_{\odot}/yr}$ and SFR$_{\textsc{disk}} = (8\pm 2)~\si{M_{\odot}/yr}$. Following the relation presented in \citet[][]{Poggianti2019a}:
\begin{equation}
    \label{LX_SFR_1}
    L_{X(0.5-10.0~\si{\kilo eV})} (\si{erg /\second})=7.6\times10^{39}\text{SFR}(\si{M_{\odot}/yr}),
\end{equation}

we calculated the corresponding X-ray luminosity, $L_{\text{X,SFR}}$, in the energy band $0.5-10.0~\si{\kilo eV}$, which are, respectively, $(6.8\pm1.5)$ and $(6.1\pm1.5)\times10^{40}~\si{erg/\second}$ for the total and disk region. \\

\begin{figure}
    \centering
    \includegraphics[scale=0.7]{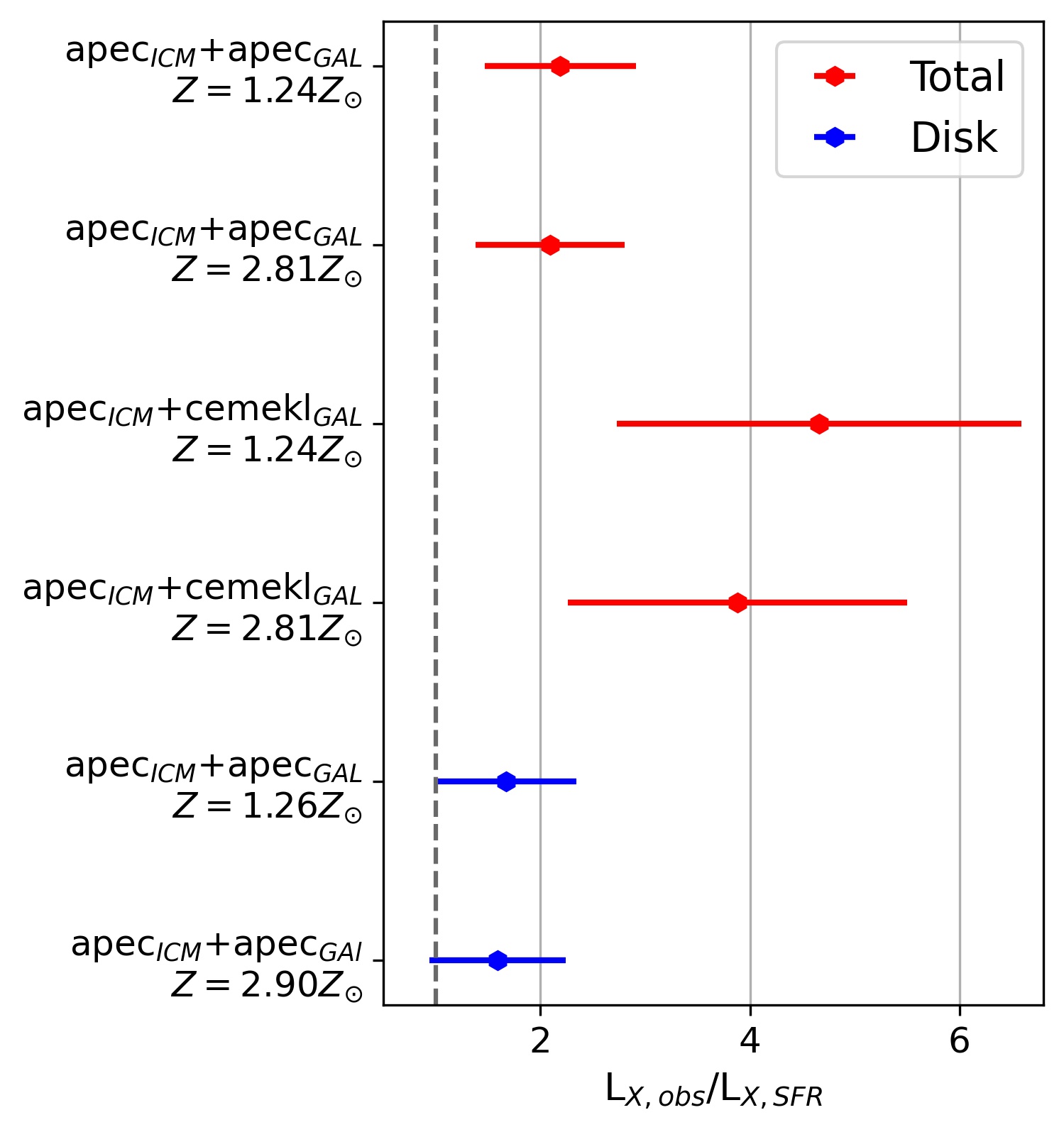}
    \caption{Ratio between the observed X-ray luminosities, $L_{\text{X,obs}}$, of the total and disk region shown in the Table \ref{tab.results}, and the X-ray luminosity associated to the SFR, $L_{\text{X,SFR}}$ (Equation \ref{LX_SFR_1}). The first four models refer to the Total region of the galaxy and the others to the Disk. The first \texttt{apec$_{\texttt{ICM}}$} represent the contribution of the ICM emission, except for the last two ones that represent the contribution of the galaxy's disk emission. The \texttt{apec$_{\texttt{GAL}}$} component describes the X-ray emission of a single temperature plasma, \texttt{cemekl$_{\texttt{GAL}}$} represents a multi-temperature plasma and \texttt{mkcflow$_{\texttt{GAL}}$} represents a multi-phase gas in radiative cooling. The dashed grey line points $L_{\text{X,obs}}$/ $L_{\text{X,SFR}}=1$.}
    \label{fig:SFR}
\end{figure}

The comparison between the observed and estimated X-ray luminosities is shown in Figure \ref{fig:SFR}. It emerges that, in the Total region, the observed X-ray luminosities are in excess with respect to what expected from the current SFR. Specifically, the excesses are by a factor of $\sim\times2$ and $\times4$ for the different spectral models. This indicates that the X-ray emission of JO194 cannot be the result of the star formation activity alone, and instead is produced by a different process. However in the disk, where most of the star formation takes place \citep[][]{Poggianti2019}, the luminosities are marginally consistent with each other ($L_{\text{X,obs}}/L_{\text{X,SFR}}=1.7\pm0.7$ and $1.6\pm0.6$, where $L_{\text{X,obs}}$ are the luminosities of the model \texttt{phabs$\cdot$(apec$_{\texttt{ICM}}$+apec$_{\texttt{GAL}}$)}). Therefore, we speculate that this additional process is more pronounced in the total galaxy than in the disk. Due to the fact that the total region actually includes the disk itself, we suggest that this additional channel to form the X-ray plasma is taking place mostly in the regions of the galaxy outside of the disk, that are the unwinding arms. \\

We note that similar X-ray luminosity excesses ($\sim10$) have been observed also in \citet[][]{Poggianti2019a} for JW100, and \citet[][]{2021ApJ...911..144C} for JO201. These discrepancy in X-ray luminosities further corroborate the idea that the star formation alone cannot provide the total X-ray luminosity, so, as already suggested in the literature, additional processes must concur in the origin of the X-ray emission in jellyfish galaxies.\\

The spectral analysis results (Table \ref{tab.results}) allows to speculate on the origin of the X-ray plasma. To begin with, our results rule out the hypothesis  that the X-ray emission could be powered by the SFR alone, either the nonthermal emission from HMXB or the thermal emission. On the other hand, the average temperature and the X-ray luminosity excess resemble the results previously reported in literature for other ram pressure stripped galaxies \citep[e.g.,][]{2010ApJ...708..946S, Zhang2013, 2019MNRAS.482.4466P, 2021ApJ...911..144C,Sun2021}. Therefore, as concluded by the previous studies, we argue that the X-ray plasma in JO194 may be the outcome of the interaction between ICM and ISM triggered by RPS. However, the nature of the interaction is poorly understood.\\
 
In this work we add a new piece to the puzzle. The X-ray spectrum of J0194’s arms is reasonably described by the model \texttt{phabs$\cdot$(apec$_{\texttt{ICM}}$+mkcflow$_{\texttt{GAL}}$)} that is the sum of the ICM thermal emission along the line-of-sight, and the emission of the hot plasma cooling onto the galaxy. Although this fitting model has a reduced $\chi^{2}$ slightly larger than the competing \texttt{phabs$\cdot$(apec$_{\texttt{ICM}}$+apec$_{\texttt{GAL}}$)}  model (see Table \ref{tab.results}), it allows the first solid metallicity measurement of the hot plasma generated by the interaction between ICM and ISM ($Z=0.24^{+0.19}_{-0.12}~\si{Z_{\odot}}$). This value is compatible within the uncertainties with the ICM abundance ($Z=0.35\pm0.07~\si{Z_{\odot}}$). This result would support the idea that the X-ray emission is produced by the radiative cooling of the metal-poor ICM onto JO194. The ICM cooling in cool-core clusters has been observed to be associated with extended H$\alpha$ emission \citep[][]{1997ApJ...486..242V}. Therefore, we tested if the H$\alpha$ luminosity expected from the ICM radiative cooling was  consistent with the upper limit imposed by the observed H$\alpha$ luminosity of JO194 ($L=(1.24 \times 10^{41}) \pm (1.12 \times 10^{37}) $ erg s$^{-1}$). The cooling H$\alpha$ luminosity is:
    \begin{equation}
        L_{H\alpha}=3.8 \times 10^{39} \dot{M}_{100}=9.1 \cdot 10^{36} \text{erg s}^{-1}
    \end{equation}
    where $\dot{M}_{100}=0.25^{0.06}_{-0.05}$ is the mass cooling rate in units of 100 solar masses per year. The resulting luminosity is negligible with respect to the observed H$\alpha$ luminosity due to star formation ($L=(5.66\pm0.01)\cdot10^{40}~\si{erg/\second}$). We conclude that the ICM radiative cooling scenario does not violate the observations, and the warm gas in JO194 is mostly photoionized by the star formation.\\
    
    However, the observed physical parameters of the ICM ($kT=4.44\pm 0.16$ keV, $n_{e}=1.39\pm 0.01~\si{cm^{-3}}$) imply that the ICM cooling time \citep[e.g.,][]{Sarazin_1986} is of the order of $\sim 30~\si{\giga yr}$, thus resulting much larger than the Hubble time. Therefore, the pure ICM radiative cooling cannot be the only origin of the X-ray emission. A previous phase of ISM-ICM mixing seems to be necessary to rise the density and lower the temperature of the ICM, which entail a decrease of its cooling time \citep[e.g.,][]{Gronke_2018, Fielding_2020}, and stimulates the radiative cooling, that results in the extended X-ray emission. Indeed \citet[][]{Franchetto2021} already reported on evidence of ISM-ICM mixing, traced by  metallicity gradients, in the stripped tails of jellyfish galaxies.
  
 \subsection{Connections between X-ray and H$\alpha$ surface brightness}
 \label{Sun}
In order to unfold the connection between X-ray emitting plasma and the warm gas in the galaxy, we studied the ratios between the X-ray ($SB_{X}$) and the H$\alpha$ ($SB_{H\alpha}$) surface brightness in the arms, the disk and the total region that represent JO194. Previous evidence of a spatial correlation between X and H$\alpha$ have been presented in \cite{Poggianti2019a, 2021ApJ...911..144C, Sun2021}. The $SB_{X}/SB_{H\alpha}$ ratio can be a signature of the physical process that produce the warm and hot gas. In case of X-ray and H$\alpha$ emission both powered by star formation, the expected $SB_X/SB_{H\alpha}$ ratio can be estimated by combining Equation \ref{LX_SFR_1} with the  SFR(H$\alpha$) relation reported in Section \ref{SFR} resulting in:
\begin{equation}
    SB_X/SB_{H\alpha}=3.48\cdot10^{-2}
    \label{kenny-pog}
\end{equation}
In the case of jellyfish galaxies, \citet[][]{Sun2021} observed that in their tails holds a different, higher ratio of:
\begin{equation}
    SB_X/SB_{H\alpha}=3.48\pm0.25
    \label{sun}
\end{equation}
which could be evidence of a different origin of the X-ray plasma with respect to the star formation, that is the interaction between the ISM and the ICM. \\

\begin{figure}
    \centering
    \includegraphics[scale=0.8]{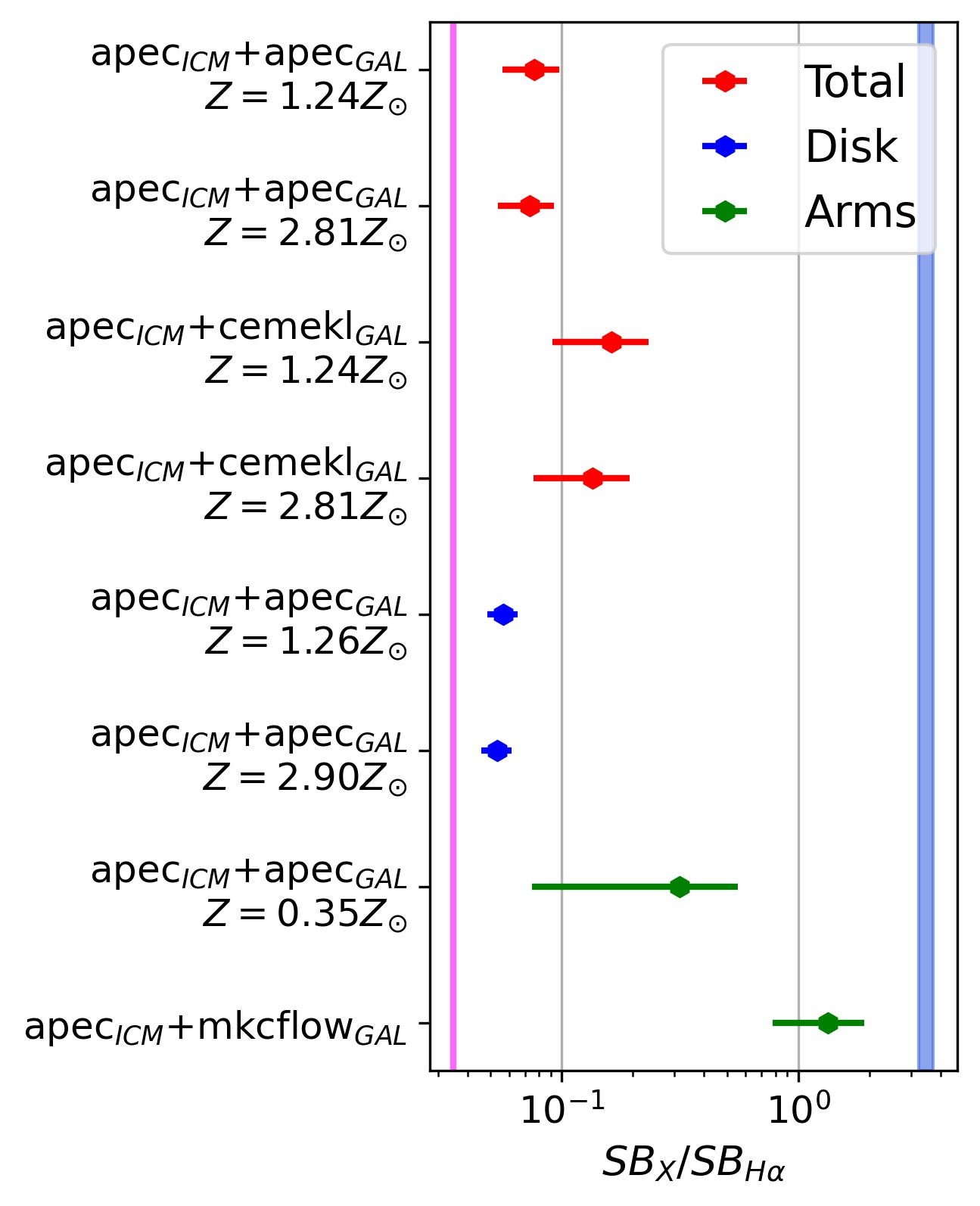}
    \caption{Total, disk and arms' surface brightness ratios. The magenta and the blue vertical ranges are respectively the ratio predicted by the SFR calibrators (Equation \ref{kenny-pog}) and the one presented in \citet[][]{Sun2021}.}
    \label{fig:grafico}
\end{figure}

In the case of JO194, we computed the average X-ray surface brightness by dividing the fluxes of the galactic component in the 0.5-10.0 keV band corrected for the absorption (see Table \ref{tab.results}) by the surface of the respective regions (Figure \ref{fig:JO194_X_estraction}). The corresponding H$\alpha$ surface brightness derived by dividing the H$\alpha$ flux measured in the Total, Disk and Arms regions from the emission-only, dust-corrected MUSE image in the spaxels with $S/R>3$ (Figure \ref{fig:ottico}) by the total area of the corresponding regions. The resulting $SB_{H\alpha}$ of total, disk and arms are, respectively, 0.9, 2.2 and $0.1\cdot 10^{-16}\si{erg/s/cm^{2}/arcsec^{2}}$. The $SB_X/SB_{H\alpha}$ are shown in Figure \ref{fig:grafico}, together with the ratios reported in Equation \ref{kenny-pog} (magenta) and \ref{sun} (blue).\\

The $SB_X/SB_{H\alpha}$ measured for JO194 are in between the values reported in in Equation \ref{kenny-pog} and \ref{sun}, with the disk and total being closer to the former, and the arms' \texttt{mkcflow$_{\texttt{GAL}}$} model being closer to the latter. Nevertheless, the $SB_X/SB_{H\alpha}$ measured for the arms is not in agreement with the ratio presented in \citet[][]{Sun2021}. This result may suggest that the X-ray emission in the disk is mainly powered by the local star formation, whereas this is not true for the arms. In this case, the fact that the observed $SB_X/SB_{H\alpha}$ ratios are lower than the values observed in jellyfish galaxies' tails (see Equation \ref{sun}) might indicate that the stripped arms are not fully evolved into a tail, at least in terms of X-ray properties. We speculate that the galaxy is still in an early stage of its evolution into a jellyfish galaxy, so in the stripped arms the ICM and ISM have not reach yet the balance observed in the other, more evolved jellyfish galaxies.

\subsection{Comparison with other GASP jellyfish galaxies} \label{comparison}
\begin{table*}[ht]
\caption{Comparison between the properties of different jellyfish galaxies: JO104, JO201, JW100. $R_{\text{cl}}$ is the cluster-centric distance and $R_{200}$ is the virial radius. The Mach number and the Ram Pressure of JW100 are lower limits (\cite{Poggianti2019a}). The properties of JO201 are shown in \cite{2017ApJ...844...49B} and \cite{2021ApJ...911..144C}.}
\label{tab.confronto}
\centering

\begin{tabular}{llll}
\toprule
Properties & JO194 & JO201 & JW100 \\
\midrule
$M_{\ast} (\si{M_{\odot}})$ & $1.6\cdot 10^{11}$ & $3.6\cdot 10^{10}$ & $3.2\cdot 10^{11}$ \\

$R_{\text{cl}}$ ($\si{\kilo pc}$) & $269$ & $360$ & $83$ \\

$R_{\text{cl}}/R_{200}$ & 0.17 & 0.18 & 0.06 \\

$kT_{\textsc{average}}(\si{keV})$ & $0.79\pm 0.03$ & $0.76\pm 0.08$ & $0.82\pm 0.11$ \\

$kT_{\text{ICM}}(\si{keV})$ & $4.44\pm 0.16$ & $7.10\pm 0.20$ & $3.50\pm 0.10$ \\

$P_{\text{ICM}}(\si{erg/c\meter^{3}})$ & $1.87\cdot 10^{-11}$ & $2.16\cdot 10^{-11}$ & $3.41 \cdot 10^{-11}$ \\

$\rho_{\text{ICM}}(\si{g/c\meter^{3}})$ & $2.63\cdot 10^{-27}$ & $1.89 \cdot 10^{-27}$ & $5.8\cdot 10^{-27}$ \\

$n_{e}$($\si{c\meter^{-3}}$) & $1.39\cdot 10^{-3}$ & $1.00\cdot 10^{-3}$ & $3.20\cdot 10^{-3}$ \\ 

Mach Number & $1.9$ & $2.4$ & $>2.0$ \\

Ram Pressure ($\si{erg/c\meter^{3}}$) & $1.08\cdot 10^{-10}$ & $2.11\cdot 10^{-10}$ & $>1.90\cdot 10^{-10}$  \\
\bottomrule
\end{tabular}
\end{table*}

We compare the results of our study with the previous studies of the GASP galaxies JO201 \citep[][]{2019MNRAS.482.4466P} and JW100 \citep[][]{Poggianti2019a} (see Table \ref{tab.confronto}). Although these galaxies are different in terms of stellar mass, velocity and environment properties, they show several similarities in their X-ray emission. These can help understanding the origin of the X-ray emitting plasma and the evolution of the galaxies subjected to ram pressure stripping.\\

Specifically, these galaxies share the following properties:
\begin{itemize}
\item The observed X-ray luminosity for the three galaxies exceeds by a factor $\sim3-10$ that expected from the current SFR inferred from the H$\alpha$ emission; 
    \item The galactic average X-ray temperatures derived using a single {\tt apec$_{\texttt{GAL}}$} model is in between $0.7$ and $1.0$ keV. This value is neither consistent with the temperature of the surrounding ICM ($4.44\pm 0.16$), nor the usual temperature of the X-ray emitting plasma in star-forming spiral galaxy \citep[$\sim0.2$ and $\sim0.65~\si{\kilo eV}$, e.g.,][]{Owen:2009}. In addition, this value is in agreement with the typical temperatures of X-ray tails of cluster late-type galaxies \citep[$\sim0.9$ keV, ][]{Sun2021}. The fact that these systems, despite different environmental conditions and evolutionary stage, have consistently developed extended X-ray plasma with similar properties suggests that the process which originated this hot plasma is both fast and stable.
    \item The Mach number of the GASP galaxies are similar: $1.9 - 2.4 $. Though the considered jellyfish galaxies are located in different environments (i.e., different ICM sound speed), this similarity implies that the galaxies are moving at supersonic velocities and are subjected to similar hydrodynamical phenomena like mixing or shock heating; 
    \item The ICM thermal pressure surrounding these galaxies ranges between 1.87 and 3.41$\times 10^{-11}$ erg cm$^{-3}$. These values are consistent with the threshold of $\sim0.9\times10^{−11} $erg cm$^{−2}$ defined in \citet[][]{Tonnesen_2011} for the formation of bright H$\alpha$ and X-ray filaments. Also the ram pressure felt by these galaxies appears to be similar ($1-2\times10^{-10}$ erg cm$^{-3}$), however these estimates are limited by the fact that the the total velocities of these galaxies are still unknown;
    \item These galaxies show extended, nonthermal radio emission \citep[][, M\"uller et al., in preparation]{Ignesti_2022}. As previously outlined in \citet[][]{Muller_2021} and \citet[][]{Ignesti_2022}, the co-existence of radio and X-ray extended emission could be explained by the ICM draping \citep[e.g.,][]{Pfrommer_2010}. By accreting hot, magnetized ICM, the magnetic field in jellyfish galaxies's tails can be amplified thus resulting in extended radio emission. In the case of JO194, the accretion and cooling of hot ICM onto the spiral arms would be consistent with the result of the spectral analysis (see Section \ref{spec_results}).
\end{itemize}

\section{Summary and conclusions}
\label{conclusion}
We analyzed an archival \textit{Chandra} observation of the cluster Abell 4059, focusing on the spectral analysis of the X-ray emission associated to jellyfish galaxy JO194. We investigated different regions of the galaxy, named Total, Disk, and Arms of JO194, where the latter indicates the Total region without the Disk. We found that the X-ray emission comes from a plasma with a temperature of $0.79^{+0.03}_{-0.04}~\si{\kilo eV}$, which is in line with the previous results reported for similar galaxies.  The spectral analysis of the arms estimate of the hot plasma metallicity, that is $Z=0.24^{+0.19}_{-0.21}~\si{Z_{\odot}}$. This value is consistent with that of the surrounding ICM, therefore we argue that the X-ray emission could be produced by the ICM that is cooling onto the cold ISM. \\

We compared the X-ray luminosities obtained by the spectral analysis of the disk and total with the X-ray luminosities associated with the current SFR, finding an excess of factor $\sim 3$ over the entire galaxy. This indicates that the star formation cannot be the only responsible for the X-ray emission of the galaxy. Moreover, we showed that the excess is predominant in the arms of the galaxy. Therefore, we suggest that the X-ray plasma pervading JO194 may be produced by the interplay between ICM and ISM, which results in the cooling of the former onto the galaxy. In order to explain the observed temperature of the X-ray plasma ($kT=0.79$ keV), we conclude that ICM cooling likely takes place initially via mixing between the ICM and ISM, which stimulates the subsequent phase of radiative cooling that, in turn, might result in the extended X-ray emission observed by {\it Chandra}.\\

In order to probe the nature of the interaction between ISM and ICM, we compared the ratio between the observed X-ray and the H$\alpha$ surface brightness with that measured in the tails of jellyfish galaxies. We found that the surface brightness ratios in the arms of JO194 are smaller than what is observed in the other ram pressure stripped galaxies. This might indicate that, at least in terms of X-ray properties and ISM-ICM balance, the arms of JO194 are still in an early stage of the evolution into a stripped tail.  \\

Finally, we compared the X-ray properties of GASP jellyfish galaxies, finding them in agreement with each other. Therefore, we suggest that the conditions required to induce extended X-ray emission in jellyfish galaxies, that is the production of hot plasma via ICM-ISM interplay (either via ICM cooling or ISM heating), are established at the beginning of the stripping, and they can persist on long time-scales, so that galaxies in different environments and evolutionary stage can present similar thermal properties.

\section*{Acknowledgements}
We thank the Referee for the suggestions that improved the presentation of the results. A.I., B.V. acknowledge the Italian PRIN-Miur 2017 (PI A. Cimatti). A.W. acknowledges financial support from ASI through the ASI-INAF agreements 2017-14-H.0. J.F. acknowledges financial support from the UNAM- DGAPA-PAPIIT IN111620 grant, México. We acknowledge funding from the INAF main-stream funding programme (PI B. Vulcani). We acknowledge financial contribution from the agreement ASI-INAF n.2017-14- H.0 (PI A. Moretti). Based on observations collected at the European Organization for Astronomical Research in the Southern Hemisphere under ESO programme 196.B-0578. This project has received funding from the European Research Council (ERC) under the European Union's Horizon 2020 research and innovation programme (grant agreement No. 833824). This research made use of Astropy, a community-developed core Python package for Astronomy \citep[][]{astropy_2013, astropy_2018}, and APLpy, an open-source plotting package for Python \citep[][]{Robitaille_2012}.  C.B. thanks Chinotto for all the constant, purrfect distractions he made during the preparation of the paper. A.I. thanks the Tokyo Ska Paradise Orchestra's music for providing the inspiration during the preparation of the draft.

\software{CIAO (v4.13; \citet[][]{2006SPIE.6270E..1VF}), Sherpa \citep[][]{2001SPIE.4477...76F, 2007ASPC..376..543D, doug_burke_2020_3944985}, XSPEC (v12.11.1; \citet[][]{1996ASPC..101...17A}), astropy \citep[][]{astropy_2013,astropy_2018}, APLpy \citep[][]{Robitaille_2012}, PYQZ \citep[][]{Dopita_2013, 10.1093/mnras/stv749}}

\bibliography{JO194_bib}{}
\bibliographystyle{aasjournal}

\end{document}